\documentclass[conference]{IEEEtran}

\IEEEoverridecommandlockouts
\usepackage{amsfonts}
\usepackage{amsmath}
\usepackage{balance}
\usepackage{multicol}

\usepackage[printonlyused]{acronym}
\usepackage{stfloats}

\usepackage{ifpdf}
\ifCLASSINFOpdf
   \usepackage[pdftex]{graphicx}
   \graphicspath{{./figures/pdf/}}
   \DeclareGraphicsExtensions{{.pdf}}
\else
   \usepackage[dvips]{graphicx}
   \graphicspath{{./figures/eps/}}
   \DeclareGraphicsExtensions{{.eps}}
\fi

\acrodef{LDPC}[LDPC]{low-density parity check}
\acrodef{BEC}[BEC]{binary erasures channel}
\acrodef{BSC}[BSC]{binary symmetric channel}
\acrodef{BBU}[BBU]{baseband unit}
\acrodef{AWGN}[AWGN]{additive white Gaussian noise}
\acrodef{SNR}[SNR]{signal-to-noise ratio}
\acrodef{SINR}[SINR]{signal-to-interference-and-noise ratio}
\acrodef{C-RAN}[C-RAN]{Centralized Radio Access Network}
\acrodef{MRS}[MRS]{Max Rate Selection}
\acrodef{EJF}[EJF]{Easiest Job First}
\acrodef{SCC}[SCC]{Scheduling with Complexity Cutoff}

\begin{document}
\abovedisplayskip=5pt
\belowdisplayskip=5pt
\abovedisplayshortskip=0pt
\belowdisplayshortskip=0pt

\newcommand{\comment}[1]{\textcolor{red}{#1}}
\newcommand{\TBD}[1]{\textcolor{red}{TBD: #1}}

\hyphenation{multi-symbol}
\title{Complexity-Aware Scheduling \\ for an LDPC Encoded C-RAN Uplink}
\author{\IEEEauthorblockN{ Kyle Whetzel and Matthew C. Valenti\\
Lane Dept. of Comp. Sci. and Elect. Eng.\\
West Virginia University\\
Morgantown, WV 26506--6109\\
Email: valenti@ieee.org} 
}

\date{}
\maketitle

\thispagestyle{empty}
\vspace{-0.80cm}

\begin{abstract}
 \ac{C-RAN}  is a new paradigm for wireless networks that centralizes the signal processing in a computing cloud, allowing commodity computational resources to be pooled.  While C-RAN improves utilization and efficiency, the computational load occasionally exceeds the available resources, creating a computational outage. This paper provides a mathematical characterization of the computational outage probability for low-density parity check (LDPC) codes, a common class of error-correcting codes.   For tractability, a binary erasures channel is assumed.  Using the concept of density evolution, the computational demand is determined for a given ensemble of codes as a function of the erasure probability.  The analysis reveals a trade-off: aggressively signaling at a high rate stresses the computing pool, while conservatively backing-off the rate can avoid computational outages. Motivated by this trade-off, an effective computationally aware scheduling algorithm is developed that balances demands for high throughput and low outage rates. \end{abstract}
\vspace{-0.4cm}

\section{Introduction} \label{Section:Intro}
Fifth-generation (5G) networks are expected to be much denser and offer higher operating frequencies than their predecessors \cite{boccardi2014five,7529132}.  The implication of densification is that there will be many more base stations with a higher variability in traffic.  As signal processing techniques used in wireless networks get more sophisticated, the digital baseband processors will continue to dominate the price, footprint, and power requirements of a wireless network.  

A recent trend in wireless networking has been to consolidate the baseband processing of multiple access points. An early version of such consolidation was distributed antenna systems \cite{saleh1987distributed}.  A more recent concept is that of \ac{C-RAN}, whereby the baseband processors of several base stations are colocated in a central processing farm \cite{checko}.  The benefits of C-RAN are numerous and stem from the sharing of commodity hardware by base stations with diverse processing needs.  Moreover, C-RAN allows for sophisticated collaborative signal processing, for instance by performing multiuser detection across base stations \cite{942518}, by transmitting distributed space-time codes \cite{944624} or by performing network MIMO \cite{7143324,7465790}.

When processing is shared by a limited pool of resources, there is a chance that the instantaneous processing demands cannot be met by the given resources.  When this happens, an outage occurs.  Such an outages is similar to an outage caused by channel impairments such as fading or interference.  In all such outages, the data will need to be retransmitted, assuming the presence of a hybrid-ARQ retransmission protocol.

In this paper, we explore the computational requirements of a typical wireless uplink.   The focus is on the transmission from the mobile unit to the base station, known as the {\em uplink}. Transmissions are assumed to be encoded with \ac{LDPC} codes, a common class of error-correcting codes.   For tractability, a \ac{BEC} is assumed.  Using the concept of \emph{density evolution}, the computational demand is determined for a given ensemble of codes as a function of the erasure probability.  The analysis reveals a trade-off: aggressively signaling at a high rate stresses the computing pool, while conservatively backing-off the rate can avoid computational outages. Motivated by this trade-off, an effective computationally aware scheduling algorithm is developed that balances demands for high throughput and low outage rates.   Simulation results show the benefits of using such a computationally aware scheduling algorithm.  

The remainder of this paper is organized as follows.  Section II gives an overview of the \ac{C-RAN} concept.  Section III reviews the salient points of \ac{LDPC} codes, providing a mathematical framework describing the complexity of \ac{LDPC} codes when used over a \ac{BEC}.  Section IV discusses how the rate-optimized \ac{LDPC} codes were identified, and shows their complexity as a function of the erasure probability.  Section V discusses the various schedulers that were considered.  Section VI provides the results of a simulation-based analysis of the schedulers.   Finally, Section VII concludes the paper and discusses potential future work.

\section{Overview of C-RAN}
The components of a cellular base station can be grouped into two main entities: The part that does analog-domain processing (e.g., power amplification and RF circuitry) and the digital \emph{baseband} processor. In a traditional cellular network, these components are packaged together in a single unit per base station. However, as technology improved, the analog components became lighter and cheaper allowing them to be placed close to the antenna at the top of towers. The digital equipment, the so called \ac{BBU}, was left at the base of towers in an equipment shack. This separation of RF and BBU has several benefits. First and foremost the antenna feed line, which is a major source of loss, can be made very short. Also the baseband equipment can be placed in a controlled enclosure with temperature control and plentiful power. 

Once analog to digital conversion (ADC) is performed atop a tower, the sampled signal can then be fed though a high speed optical cable to the BBU. This link between the RF component and BBU is known as the {\em fronthaul} link. Common Public Radio Interface (CPRI) is the protocol used for this link. CPRI transmits a constant bit rate up to 12 Gbps. It allows for a link of up to 40 kilometers with a latency of only 0.1 ms. This distance allows BBUs for multiple cells to be consolidated in a so called {\em baseband hotel}. This lends to cost effectiveness because less infrastructure is needed for delivering power, temperature control, and network links to the BBUs \cite{checko}. 
These baseband hotels also allow for joint processing and the sharing of computing resources. This sharing of resources creates a statistical multiplexing gain. The gain comes from exploiting temporal and spatial traffic fluctuations which are inherent to cellular networks. 

Mobile networks are hard real-time systems with tight timing and protocol constraints \cite{GC1}. In a C-RAN, sampled data must be quickly transmitted over the fronthaul and processed in real time with a hard deadline. It is possible for the capacity of the fronthaul link to not be sufficient to meet these deadlines. There is also finite computing resources at each baseband hotel. It is possible the computational load demanded by the network is greater than the processing resources available. If either of these situations arise, a {\em computational outage} is said to occur. These computational outages are just as detrimental to wireless systems as outages caused by fading and interference \cite{transactions}. 

Of the tasks handled at BBUs, forward error control (FEC) is amongst one of the most computationally intensive \cite{GC2}. FEC aims to add redundancy to transmitted data such that it can recover from errors occurred during transmission. By adding this redundancy data rate is compromised.  A lot of literature has focused on maximizing the code rate to approach the Shannon limit. However, signaling at higher data rates requires more computational resources for decoding. This trade-off becomes very important in C-RAN systems. It is the objective of a C-RAN {\em scheduler} to select the coding scheme in real time to best meet the demands of the system.

\section{The Complexity of LDPC Codes} 
\ac{LDPC} codes use an iterative decoder that is defined on a {\em Tanner Graph} \cite{book:LDPC}.  As such, the decoding complexity depends on the number of decoding iterations and the layout of the graph.  The number of decoder iterations can be predicted using a concept known as {\em density evolution}.  For ease of exposition, we focus on density evolution for the \ac{BEC} in this paper, though the work could be extended to other channels, including \ac{BSC} and \ac{AWGN} channels.  

In the \ac{BEC}, the input may be a data 0 or a data 1, while the output may be a data 0, data 1, or an erasure.  With probability $\epsilon$, called the \emph{erasure probability}, the bit is erased, in which case the channel outputs an erasure, while with probability $1-\epsilon$ the bit is received correctly.  When the erasure probability is $\epsilon$, the Shannon capacity is $1-\epsilon$, meaning that a code exists that enables reliable communication at a rate very close to $R=1-\epsilon$. 

The Shannon capacity of the erasure channel can be reached by using LDPC codes.  The input to an LDPC encoder is a {\em message} $\mathbf{u}$ of length $k$ bits.  The encoder maps each message $u$ to a {\em codeword} $\mathbf{c}$ of length $n$ bits, where $n>k$. The ratio of message bits to the total number of bits in a codeword, $R=k/n$, is called the {\em code rate}. The code is denoted by $\mathcal C$, which is the set of all codewords.  

\subsection{Describing LDPC Codes}
An LDPC code may be described by its {\em parity-check matrix}, $ H$, which is a rank $(n-k)$  matrix whose rows span the dual code $\mathcal C^{\perp}$, which is the set of all length-n binary vectors that are orthogonal to every vector $c \in \mathcal C$.   The parity check matrix $H$ is  usually is full rank, in which case $H$ has $n-k$ rows. Because the rows of $H$ are from $\mathcal C^{\perp}$, and hence are orthogonal to every codeword, it follows that $c H^T  =   0$ for all $c \in \mathcal C$.   An \ac{LDPC} code is characterized by a sparse parity-check matrix; i.e., a large matrix containing very few ones. The sparseness results in a low decoder complexity, even as the code length becomes long (as is necessary to approach the Shannon capacity).

 An example parity-check matrix is as follows:
\begin{eqnarray}
   H  & = &
   {\left[
     \begin{matrix}
       1 & 1 & 1 & 0 & 0 & 0 & 0 & 0 & 0 \\
       0 & 0 & 0 & 1 & 1 & 1 & 0 & 0 & 0 \\
       1 & 0 & 0 & 1 & 0 & 0 & 1 & 0 & 0 \\
       0 & 1 & 0 & 0 & 1 & 0 & 0 & 1 & 0 \\
       0 & 0 & 1 & 0 & 0 & 1 & 0 & 0 & 1
     \end{matrix}
   \right]}.
   \label{H_matrix}
\end{eqnarray}
By expanding $c H^T=0$, a set of parity-check equations may be found, one for each row of $H$.  For instance, the parity-check equations for $H$ matrix of (\ref{H_matrix}) are:
\begin{eqnarray*}
   c_0 \oplus c_1 \oplus c_2 & = & 0 \\
   c_3 \oplus c_4 \oplus c_5 & = & 0 \\
   c_0 \oplus c_3 \oplus c_6 & = & 0 \\
   c_1 \oplus c_4 \oplus c_7 & = & 0 \\
   c_2 \oplus c_5 \oplus c_8 & = & 0.
\end{eqnarray*}
Code bit $c_j$ is said to \emph{participate} in the $i^{th}$ parity-check equation if there is a 1 in the $(i,j)^\mathsf{th}$ position of $H$. 

\begin{figure}[t]
     \centering
     \includegraphics[width=.5\textwidth]{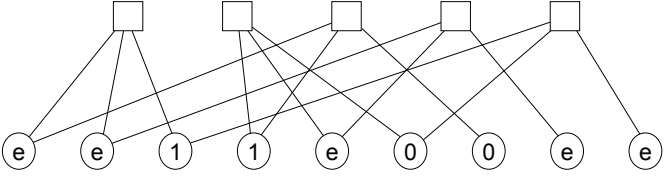}
     \caption{Example Tanner Graph}
     \label{fig:TG}
     \vspace{-0.25cm}
\end{figure}

A \emph{Tanner graph} is a bipartite graph that represents an $H$ matrix. In Tanner Graphs, the two independent sets of vertices are called the \emph{check} nodes and the \emph{variable} nodes. The check nodes represent the $n-k$ parity-check equations. The variable nodes represent the $n$ code bits. If an entry in the parity-check matrix $H_{i,j}=1$ then the $i^{th}$ check node is connected to the $j^{th}$ variable node.  Thus, the Tanner graph is a pictorial representation showing which code bits participate in which parity-check equations.  Fig. \ref{fig:TG} shows the Tanner Graph for the parity check matrix given in (\ref{H_matrix}).

Decoding can be performed using the Tanner Graph. First the variable nodes are loaded with the observed code bits. In  Fig. \ref{fig:TG} an example received message containing erasures is shown. Each check node examines its incident edges to see if it is connected to a single variable node containing an erasure. If this is the case those check nodes can then correct that erasure using the logic of a single parity-check code.   This process of checking incident edges at each check node is then iterated until all erasures are corrected or there are no more possible corrections to be made. 

The {\em degree} of a check node is the number of incident edges connected to it. This number is equal to the  the number of ones, or the {\em Hamming Weight}, of the corresponding row in the $H$ matrix. The complexity of decoding using a parity-check matrix is dependent on the degree of the check nodes. Therefore, it is desired to have $H$ matrices with low row weight. 


If the rows of a LDPC parity-check matrix have constant weight, or constant check-node degree $d_c$, the code is said to be {\em check regular}. If the Hamming weight of the columns  of a LDPC parity-check matrix are also constant, that is the variable-node degree $d_v$ is constant, the code is said to be {\em regular}.  For brevity, the shorthand ($d_v,d_c$) is used to specify a regular code with variable-node degree $d_c$ and check-node degree $d_c$. 

Although  regular codes are simple and perform moderately well, they are not capable of achieving capacity. Irregular LDPC codes are those without a constant degree, and when properly designed are capable of  achieving capacity. When designing irregular LDPC codes, the variable node distribution is not constant. The check node degree is however typically held constant, or close to it. 

The design of irregular LDPC codes consists of choosing the proper degree distribution.  Let $\rho_i$ denote the fraction of {\em edges} touching degree $i$ check nodes, and $\lambda_i$ denote the fraction of {\em edges} touching degree $i$ variable nodes. The degree distributions can be described in polynomial form. $\rho(x) = \sum_i \rho_i x^{i-1}$ is the distribution for check nodes, and $\lambda(x) = \sum_i \lambda_i x^{i-1}$ for variable nodes.  From \cite{book:LDPC} , the rate of the code can be found in terms of the polynomial form of the degree distributions as 
\begin{eqnarray}
R
& = &
1 - 
\frac{\int_0^1 \rho(x)dx}{\int_0^1 \lambda(x)dx}. 
\end{eqnarray}

\subsection{Density Evolution}\label{sec:DE}
Density evolution is a tool for predicting the erasure probability as a function of the number of decoder iterations.  For a LDPC code over a BEC, \cite{book:LDPC}  states the probability that a variable-node remains erased after the $\ell^{th}$ iteration is
\begin{eqnarray}\label{eq:DE}
   \epsilon_\ell
   & = &
   \epsilon_0
   \lambda
   \left(
      1 -
      \rho
      \left(
         1 - \epsilon_{\ell-1}
      \right)
   \right)
\end{eqnarray}
where $\epsilon_0$ is the initial erasure probability before any error correcting has been performed. A code is said to {\em converge} if the erasure probability is reduced to or below a certain threshold after a sufficiently large number of iterations.  
Let the \emph{convergence threshold} be denoted by $\epsilon_\mathsf{thresh}$. The criteria for convergence is that the erasure probability is constantly decreasing; i.e., if $\epsilon_{\ell} < \epsilon_{\ell-1}$ for all $\ell$ in which $\epsilon_{\ell}>\epsilon_\mathsf{thresh}$. 

\begin{figure}
     \centering
          \includegraphics[width=.45\textwidth]{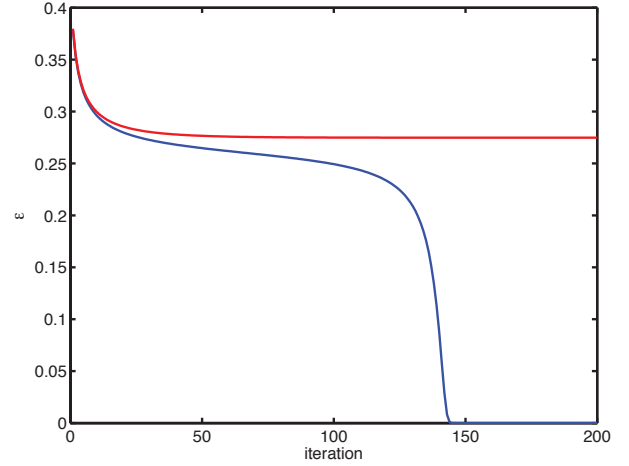}          
        \caption{Density Evolution of a (3,6) regular code. The threshold $\epsilon^*=0.4294$ for this code. One plot is for $\epsilon_0=0.43$ the other $\epsilon_0=0.429$  }\label{fig:DE*}         
\end{figure}

From the requirement that $\epsilon_{\ell} < \epsilon_{\ell-1}$ and (\ref{eq:DE}), 
\begin{eqnarray}\label{eq:DE2}
   \underbrace{
   \epsilon_0
   \lambda
   \left(
      1 -
      \rho
      \left(
         1 - \epsilon_{\ell-1}
      \right)
   \right)}_{\epsilon_\ell}
   & < &
   \epsilon_{\ell-1}.
\end{eqnarray}
By performing the change of variable $\epsilon_{\ell-1} \rightarrow x$ and defining the function
\begin{eqnarray}\label{eq:DE3}
  f( \epsilon_0, x )
   & = &
   \epsilon_0
   \lambda
   \left(
      1 -
      \rho
      \left(
         1 - x
      \right)
   \right),
\end{eqnarray}
it is found that decoder convergence is possible if and only if $f( \epsilon_0, x) < x$ for all $\epsilon_\mathsf{thresh}\leq x \leq \epsilon_0$. It is useful to find the largest $\epsilon_0$ for which the code converges.  This initial erasure probability threshold, will be denoted as $\epsilon^*$. Fig. \ref{fig:DE*} shows density evolution for a code when the $\epsilon_0$ is just above and just below $\epsilon^*$

Rearranging (\ref{eq:DE3}) and using $f( \epsilon_0, x) < x$ yields  
\begin{eqnarray}
   \epsilon(x)
   & = &
   \frac{x}{   \lambda
   \left(
      1 -
      \rho
      \left(
         1 - x
      \right)
   \right) }.
\end{eqnarray}
To find $\epsilon^*$ the largest value of $\epsilon_0$ such that $\epsilon_0 < \epsilon(x)$ for all $x$ must be determined. The function $\epsilon(x)$ is a convex function, $\epsilon^*$ is its minimum value, as follows: 
\begin{eqnarray}\label{eq:estar}
   \epsilon^*
   & = &
   \min \{ \epsilon(x): \epsilon(x) >x\}.
\end{eqnarray}

\subsection{LDPC Code Complexity}
The complexity of a decoder is the amount of processing resources required to successfully decode a received message. Let $\mathrm{K}$ be complexity requried per codeword, which is equal to the number of decoder iterations multiplied by the number of edges in a Tanner graph. Assuming a check-regular LDPC code $\mathrm{K}=\ell d_c (n-k)$. 

To get a better understanding of complexity versus throughput of information it would be beneficial to know the required work per data bit. Let $\mathrm{C}$ be complexity per data bit. To find $\mathrm{C}$, $\mathrm{K}$ will be normalized by dividing by the number of data bits, $k$; i.e., 
$\mathrm{C}=\frac{{\ell} d_c(n-k)}{k}$.   Knowing that $R=k/n$ the complexity per code bit can be written in terms of the code rate as
\begin{eqnarray}\label{eq:C}
   \mathrm{C}=\frac{{\ell} d_c(1-R)}{R}.
\end{eqnarray}

\section{Rate Optimized Codes}
A palette of codes is required for use by the complexity-aware scheduler.  The codes were designed by constraining them to be check-regular with $d_c=7$ and then finding the variable-node degree $\lambda(x)$ that maximizes $\epsilon^*$.  For the code to be valid it must meet the following constraints: 
\begin{eqnarray}
0 \leq \lambda_i \leq 1 & & \mbox{for $1\leq i \leq d_{max}$}\label{eq:dv} \\
\sum_{i=1}^{d_\mathsf{max}} \lambda_i  & = & 1 \\
\sum_{i=1}^{d_\mathsf{max}} \frac{\lambda_i}{i}
& = & 
\frac{1}{(d_c)(1-R)}
\end{eqnarray}
where $d_\mathsf{max}$ is the largest allowable variable node degree.  It was found that performance is enhanced by avoiding degree-1 variable nodes, which is accomplished by changing (\ref{eq:dv}) to $\lambda_1  =  0 $ and $  0 \leq \lambda_i \leq 1 \; \; \; \mbox{for $2\leq i \leq d_{max}$}$. 

With $d_\mathsf{max}= 200$, eight codes were developed with rates. Let $(R,\epsilon^*)$ denote the rate / erasure threshold pair for each code.  The developed codes had $(R,\epsilon^*)$ pairs
$(1/5,0.728)$,
$(1/4,0.708)$,
$(1/3,0.657)$,
$(2/5,0.589)$,
$(1/2,0.478)$,
$(3/5,0.367)$,
$(2/3,0.274)$, and
$(3/4,0.167)$.
Thus, the largest erasure probability that can be handled by the codes is 0.728. Any situation where the initial erasure probability is greater will result in failure to decode the message.

\begin{figure}
     \includegraphics[width=.5\textwidth]{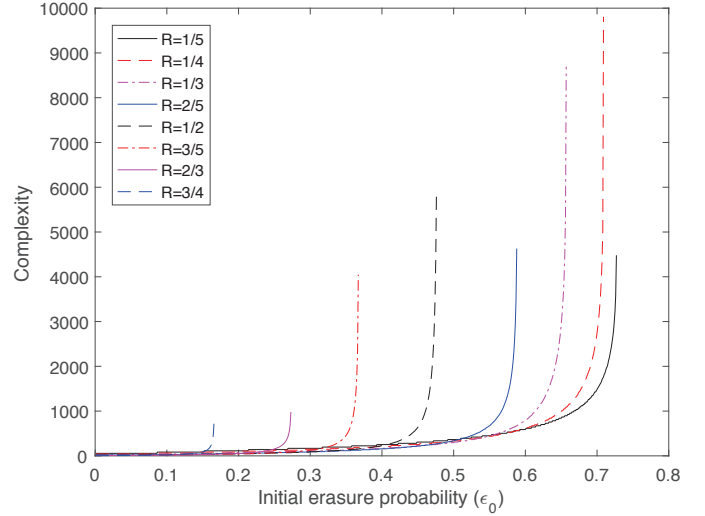}
     \centering
     \caption{Complexity Required For Convergence. Note the spike of complexity values near each codes $\epsilon^*$.}
     \label{fig:CvsE0}
\end{figure}

By using (\ref{eq:C}), complexity is calculated for various values of $\epsilon_0$ using density evolution described in section \ref{sec:DE}. The convergence threshold was set to $\epsilon_{thresh}=10^{-3}$ and the max number of iterations was set to $10^3$. Initial erasure probability was varied from 0 to just above the threshold $\epsilon^*$ for each code. The resulting complexities are shown in Fig. \ref{fig:CvsE0}. As the initial erasure probability $\epsilon_0$ approaches the threshold $\epsilon^*$ for each code, the computational complexity spikes up. This shows the desirability of complexity aware scheduling systems. If operating near $\epsilon^*$ the computation load can be lowered dramatically by switching to a lower coding rate. 


\section{Complexity Aware Scheduling for C-RAN}

Four different scheduling algorithms are considered for the C-RAN system. These include complexity unaware and complexity aware methods. Let  $\mathrm{C}_\mathsf{server}$ be the total computational complexity resources available to a computing cluster of size $N$. Let  $\mathrm{C}_{\kappa}$ be the computational complexity required at a base station  for user $\kappa$ within the computing cluster. The rate, $r_{\kappa} \in \mathcal{R}$, is the code rate assigned to the user $\kappa$. Throughput, $T$, will be defined as the average of all assigned code rates $r_{\kappa}$ contained in the computing cluster.

In the {\bf \ac{MRS}} algorithm, each user in the cluster $\kappa$ is assigned the maximum code rate available based on $\epsilon_0$ at the base station and $\epsilon^*$ of the codes in the pool. If decoding can't be performed for a certain user, $r_{\kappa}$ is set to 0. Complexity, $\mathrm{C}_{\kappa}$, is then calculated  for each user within the computing cluster as a batch process. If 
$\sum_{\kappa =1}^N\mathrm{C}_{\kappa} > \mathrm{C}_\mathsf{server}$ a computational outage occurs. The rate is set to zero for all users resulting in zero throughput. 

The  {\bf \ac{EJF}} algorithm initializes by performing the MRS code selection scheme. If 
$\sum_{\kappa =1}^{N}\mathrm{C}_{\kappa} > \mathrm{C}_\mathsf{server}$ the scheduler selects the user $\kappa$ that has the lowest complexity value, or the easiest job.  If this minimum complexity $\mathrm{C}_{ej}< \mathrm{C}_\mathsf{server}$ decoding is performed for that user and $\mathrm{C}_{ej}$ is then added to the sum of complexities already decoded, $\mathrm{C}_{sum}$. $\mathrm{C}_{sum}$ is set to 0 before any decoding has occurred. This process then repeats by selecting the easiest job of those remaining to be decoded as long as $\mathrm{C}_{sum}< \mathrm{C}_\mathsf{server}$. 

The {\bf Local Limit} algorithm starts by selecting the highest possible code rate for each user as in the MRS algorithm. It sets a limit, $\mathrm{C}_{loc}$,  on the computational resources allocated to one user. If any user $\kappa$ demands a decoder complexity greater than $\mathrm{C}_{loc}$, $r_{\kappa}$ is bumped down to a lower rate code if one exists. Complexity is then recalculated. This process is then reiterated until $\sum_{\kappa =1}^{N}\mathrm{C}_{\kappa} < \mathrm{C}_\mathsf{server}$, or there are no lower rate codes to recede to. If the case is the latter, and no more codes are available, a computational outage has occurred. 

The {\bf \ac{SCC}} algorithm also initializes by setting rates to the maximum possible as in the MRS scheme. It then checks if 
$\sum_{\kappa =1}^{N}\mathrm{C}_{\kappa} > \mathrm{C}_\mathsf{server}$. If that is the case user $\kappa^*$ is chosen such that $ \mathrm{C}_{\kappa^*}$ is the maximum $\mathrm{C}_{\kappa}$. The code rate for user $\kappa^*$, $r_{\kappa^*}$, is then decreased to the next lower coding rate if one exists. If a lower rate code does not exist for $\kappa^*$ the user $\kappa$ with the next highest $\mathrm{C}$ is then chosen. $\sum_{\kappa =1}^{N}\mathrm{C}_{\kappa}$ is then recalculated with the new coding rate. If this sum is less than $\mathrm{C}_\mathsf{server}$ decoding commences, else the process is recursed such that an updated $\kappa^*$ is selected. The recursion proceeds until the complexity constraint is met or no lower rate codes are available. 

\section{Simulation Results}
To illustrate the efficacy of the scheduling algorithms, a simulation study was conducted. In the simulations, 110 base stations are placed in a hexagonal grid pattern.  Mobile units (MU) are randomly dropped into the system such that there is at most only one active mobile per base station cell.   The number of cells with users is the channel {\em utilization}.  The  computing cluster will be a central group of adjacent cells which have pooled computing resources.

%
%

Let $\{X_i\}$ indicate the set of placed MU's and their locations.  A given MU $X_i$ is connected to a particular base station, denoted $Y_i$.  After each set of MU's are placed, the \ac{SINR} is calculated for each base station in the computing cluster. The SINR at basestation $Y_j$ is
\begin{eqnarray}\label{eq:SIR}
\gamma_j= \frac{|Y_j-X_j|^{\alpha (s-1)}}
{\Gamma^{-1}+ \sum_{i\neq j} |Y_j-X_i|^{-\alpha}|Y_i-X_i|^{s\alpha}}
\end{eqnarray}
where by $|Y_j-X_i|$ being the distance from a base station $Y_j$ MU $X_i$, $\alpha$ is the path-loss exponent, $s$ is the compensation factor for fractional power control, $\Gamma$ is the SNR at unit distance, and the summation is over all interferers. The power control factor is set to $s=0.1$, which is the value that maximizes throughput \cite{paper:s}.

Once the SINR is calculated for each base station in the computing cluster, the corresponding erasure probability is found by using the complementary cumulative distribution function (CCDF) of the SINRs obtained. The CCDF makes use of the SINRs obtained with all of the default simulation values. By definition, the CDF of a random variable X evaluated at x is the probability that X is less than x. That is $F_X(x)=P(X<x)$. Since this is a probability, as is $\epsilon_0$, the value ranges from 0 to 1. Because high SINR values need to be mapped to low initial erasure probability $\epsilon_0$ values, the CCDF is used. The complementary cumulative distribution function CCDF=1-CDF.  An exponential curve is then fitted to the CCDF. 




Unless otherwise specified, the simulations used the default values of the simulation parameters are as follows: path-loss exponent $\alpha=3$, utilization $u=100$ \%, cluster size $N=7$, SNR at unit distance $\Gamma = 20$ dB. The number of Monte Carlo trials is set to $10^5$.

\begin{figure}
     \includegraphics[width=.5\textwidth]{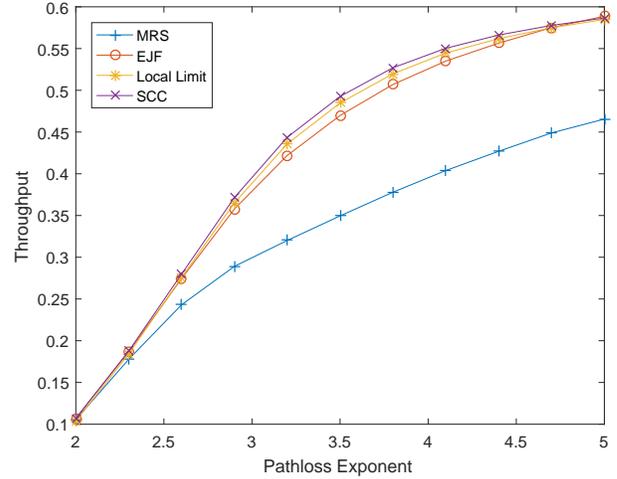}
     \centering
     \caption{Throughput versus Path-loss exponent $\alpha$. $\alpha$ is varied from a value for free space, 2, to 5 which corresponds to very rough terrain or high blockage. The computing cluster size is held constant at 7 and the channel utilization constant at 100\%. }
     \label{fig:TvsA}
\end{figure}

The first simulation explores the impact of path-loss exponent $\alpha$ as the independent variable. It was varied such that $2\leq \alpha \leq 5$, with $\alpha = 2$ corresponding to free-space and larger path-losses corresponding to rough terrain or blockage.   Fig. \ref{fig:TvsA} shows the effect of varying the pathloss exponent $\alpha$ on throughput for the different scheduling algorithms. When $\alpha=2$, the throughput for all schedulers is very low due to the high interference, which does not get sufficiently attenuated in free space.

\begin{figure}
     \includegraphics[width=.5\textwidth]{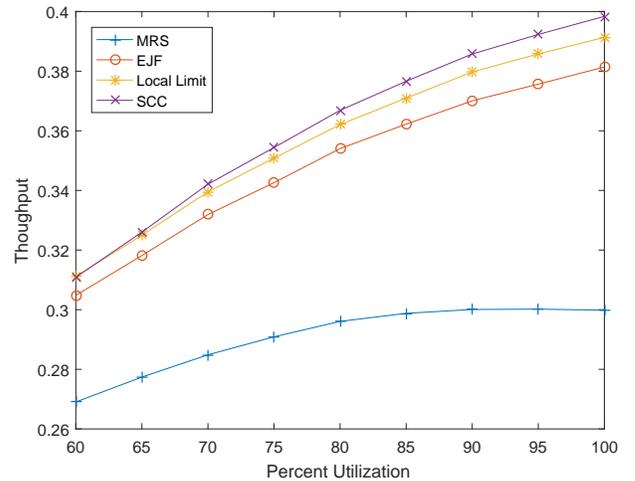}
     \centering
     \caption{Throughput versus Channel Utilization. If there is no user in a computing cluster cell the throughput for that cell is set to zero. Pathloss exponent $\alpha$ is held constant at 3 and the computing cluster size at 7.}
     \label{fig:TvsU}
\end{figure}

The second simulation explored the effect of channel utilization. The channel usage was varied between $0.6 \leq u \leq 1$.  Fig. \ref{fig:TvsU} displays the throughput when varying the channel utilization. When randomly placing the mobile units there is no guarantee a user will be placed in every cell of the computing cluster. The throughput is lower at lower channel utilization's because no data is being transmitted if there isn't a user in that cell. 

\begin{figure}
    \includegraphics[width=.5\textwidth]{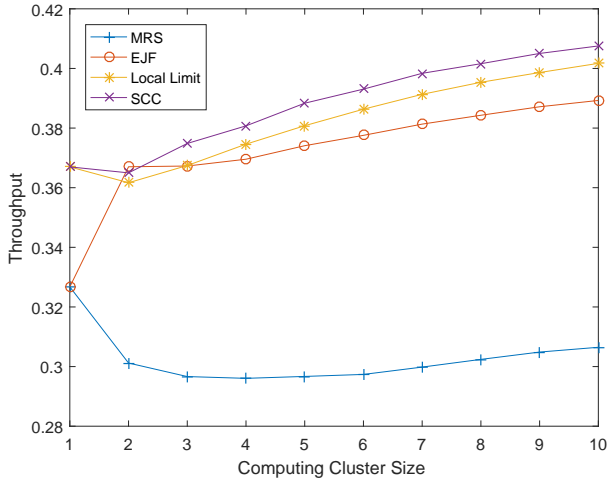}
     \centering
     \caption{Throughput versus the Computing Cluster Size. The pathloss exponent is constant at $\alpha=3$ and the channel utilization is held constant at 100\%.}
     \label{fig:TvsCS}
\end{figure}

The last simulation explored the effect of the computing cluster size, by varying in the range $1 \leq N \leq 10$.  Fig. \ref{fig:TvsCS} shows the throughput when varying the computing cluster size. The first data point here is when cluster size is set to one. This is the same as a traditional cellular network where no processing resources are shared. From the plot you can see that increasing cluster size increases the throughput for all complexity aware algorithms. 

Looking at all the simulations it can be seen that the complexity aware schemes always outperform the complexity unaware MRS. In terms of throughput, the more sophisticated SCC does as good or better for all values of the variables being varied. 

\section{Conclusions}

Advancements in technology has made it feasible to process baseband signals of wireless networks in a central processing pool. This network paradigm shift, known as C-RAN, creates many benefits. These include the ability to share computing resources among multiple cells and centralizing the management of system infrastructure. These benefits come along with some challenges. The primary being the load imposed on the front haul connection, and the risk that the demand on computing resources exceeds its capacity. When either the fronthaul capacity isn't sufficient, or the processing center doesn't have sufficient computation resources a outage occurs. From a provider's perspective this outage is no different from a more traditional impairment such as fading or interference.  

This paper has provided an analysis of the computational requirements of a C-RAN uplink, and has developed and analyzed computationally aware scheduling algorithms for C-RAN.  In order to perform a mathematically precise and tractable analysis, we have assumed that the transmissions are encoded with LDPC codes and sent over a BEC channel.  While this allows for the per-bit complexity to be accurately predicted, future work should extend the analysis to other kinds of channels. 

%

\balance

Another possible future direction is to create LDPC codes that are themselves complexity aware; i.e., designed with an additional constraint that attempts to reduce the complexity rather than just minimize the convergence threshold.  Finally, other, more sophisticated complexity aware algorithms could be developed.

\bibliographystyle{ieeetr}
\bibliography{ciss}

\end{document}